\documentclass[twocolumn,twoside, ]{article}
\usepackage[utf8]{inputenc}
\usepackage{graphicx}
\newcommand{\h}{\linebreak \hspace*{3ex}}
\newcommand{\hb}{\\ \hspace*{2ex}}

\begin{document}
\title{ON THE POSSIBILITY FOR MEASURING THE HUBBLE CONSTANT FROM
OPTICAL-TO-NIR VARIABILITY TIME DELAY IN  AGNS}
\author{V.L.\,Oknyanskij \\[2mm]
Sternberg Astronomical Institute,\hb
       Universitetskij Prospect 13,
       119899 Moscow, Russia,
       {\em  oknyan@sai.msu.ru}\\[2mm]
}
\date{}
\maketitle

ABSTRACT.
The  Optical-to-Near-infrared variability time delay have already been
reported  for a small number ($\sim 7$) of AGNs and has been firmly
established only for 5 of them. The  time delay is probably increasing
with the IR wavelengths. The most naturally this time delay can be
interpreted by the model where IR emission is attributed to
circumnuclear dust heated by the nuclear radiation.  In given model a
suggestion on narrowness of the near-infrared (NIR)  emission region is
quite natural, as far as the dust can be not saved on distances from the
nucleus  closer then some critical value, on which it is reached the
sublimation temperature  for graphite particles (Barvainis, 1987).
For  NGC 4151 case it has been shown that the  NIR region has a form of
thin ring or torus. The radius of this ring  correlates with level of the nucleus 
activity (Oknyanskij et al. 1999). This dependency of radius of the NIR emission
region from luminosity reveals itself as under object variability (as in the case
of NGC4151),  and also  when  objects with high and low luminosity are
considered.   We assume that the observed time delays allow us to derive
a redshift independent luminosity distances to AGNs and estimate a Hubble
constant.

Some problems of using this strategy for  the Hubble constant
determination are discussed.\\[1mm]

{\bf 1. Introduction}\\[1mm]
$H_0$ - the Hubble constant is fundamental parameter in standard cosmology, measuring the rate at which
the Universe is expanding.  $H_0$ connected with many other significant values in cosmology and first of
al age of Universe and distance scale. The value of the Hubble constant is still subject of intensive 
discussions in many publications.

There are 2 different groups of methods for the Hubble constant determinations
can be noted:

1. Traditional or {\it ``non direct"} methods, which use some directly measured distance
in our Galaxy, for example till Hiades, then  extrapolate it some way till
other galaxies. In this group of methods are well known:

(i) using period-luminosity dependence for variable stars (a recent successful example of this
is HST Key project (Freedman et al., 1994));

(ii) using the principle that a sample of nearby spirals of specific Hubble
type represents a ``fair" sample of intrinsic population (Sandage 1996; Goodwin, 1997);

and some others.

Current estimates of $H_0$ using methods from this group are in range $\sim 60-90~km s^{-1} Mpc^{-1}$

2. {\it ``Direct methods"}:

(i) Using Sunyaev-Zel'dovich effect (Syunyaev \&  Zeldovich 1980);

(ii) Using gravitationally lensed QSOs (Refsdal 1964);

(iii)  Time delay between variability of AGN in different UV-Optical-NIR wavelengths (Collier et
al., 1999);

(iv) and some others, for example,  using motions and line-of-sight
accelerations of water maser  emission (Miyoshi et al. 1995).

Current estimates of $H_0$ using methods from this group are in range
$\sim 30-80~ km~s^{-1}~Mpc^{-1}$.

``Direct'' methods give systematically smaller values for $H_0$ than ``non direct''.  Meanwhile these 'direct'
methods  are more model dependent.

We propose here a new  method that utilizes  the redshift-independent luminosities of AGNs   obtained 
from observed optical-to-near IR time delay.

In  chapter 2 we discuss the theory of the method.
In chapter 3 we present our published and new results of the optical-to-NIR time delay determinations
for several AGNs NGC4151 (Oknyanskij 1993, Oknyanskij et al. 1999),  QSO PQ Comae and NGC7469
and combine them with other published results on the optical-to-NIR time delays
in several other AGNs. Then we apply  our method  for determination of $H_0$
using the observation results.\\[4mm]

{\bf 2. Theory}\\[1mm]

{\it 2.1. First step idea of the method}\\[1mm]
{\it 2.1.1. Basic assumptions}\\[1mm]
(i) NIR emission is attributed to circumnuclear dust heated by 
the nuclear radiation.

(ii) The dust is spherically symmetric and smoothly distributed.

(iii) NIR emission region has a form of a smooth spherical shell.

(iii)      The dust (graphite grains) can be not survived on distances from the
            nucleus  closer then some critical value, on which it is reached the
            sublimation temperature  for graphite particles (Barvainis, 1987,  next times 
            here B1)   
   
(iv) The time delay between UV (optical)  and  NIR variations caused by simple light travel time 
effects.
 
    This critical distance, "evaporation radius" is given by  (following to B1):
\begin{equation}
r_{evap}=1.3~L_{UV,4}^{0.5}~T_{1500}^{-2.8} pc
\end{equation}
where T is the grain evaporation temperature in units of 1500 K and L is 
ultraviolet luminosity in units $10^{46} ergs~ s^{-1}$ and $r_{evap}$ is the radius in
parsecs.\\[2mm] 

{\it 2.1.2. Core of the idea}\\[1mm]
From the observations we can get the time 
delay between UV (or optical) and near IR variations in some 
AGN, which give us $r_{evap}$ and estimation of $L_{UV}^{\ast}$. Then we can use
observed flux in UV to get independent  from z distance to the object 
and estimate the $H_0$.  If we have already got the estimation of the $L_{UV,H=50}$
for  $H_0=50~km~s^{-1}~Mpc^{-1}$ then we can get estimation\\
\begin{equation}
H_0=50~{(L_{UV,H=50} / L_{UV}^\ast)}^{0.5}~km~s^{-1}~Mpc^{-1}
\end{equation}\\[2mm]

{\it 2.1.3 Problems}\\[1mm]

1. From the observations we have found that the NIR emission region 
should have form of thin ring or torus, but not spherical shell (Oknyanskij, 1999).

2. If the grains are depleted when the UV luminosity peaks, and 
cannot reform, then a dust-free hole surrounding the central source 
will be created with radius corresponding to the sublimation distance 
at the UV peak. This hole can be a problem in explanation of NIR 
variability.

3. The nature of the grain is unknown.  The evaporation temperature 
can be significantly higher then 1500~K considered in B1 and
probably can reach 2000~K  (Sanders et al 1989). The size of the
grains also can be bigger then 0.05 $\mu$  used for deriving  (1) .\\[2mm]

{\it 2.2. Next step model}\\[1mm]
Barvanis (1992) has considered  the ``survival" and ``reformation" models. The reason for this is that clouds might serve to
either protect the grain from sublimation, allowing them to serve when 
the UV flux high, or provide a medium in which grains can reform. So 
the model thus assumes that dust is located into clouds.  

For next step we can use small improvements:
we will assume that 

(i) dust is clumped into clouds with UV optical depth $\tau_{UV} \ge$1;

(ii) the dust region geometry has disklike form.

Thus model assumes clouds existing at radii well inside the sublimation 
radius  for the peak UV flux given by (1).

In place of (1) we will use here improvement of it with 2 additional 
parameters given by Sitko et al. (1993):
\begin{equation}
r_{evap}=~9 \times 10^{8}~L_{UV,46}^{0.5}~T^{-2.8}~[0.05/A_{\mu}]^{0.5}~ e^{-\tau/2} pc
\end{equation}
where  $A_{\mu}$  is graphite grain size in $\mu$, $\tau$ - is optical depth of the clouds
in UV. We will use following to Sitko et al. the same values of
parameters: $T=1700$ K, $A_{\mu}=0.15$, $\tau=1$.\\[5mm]

{\small
\begin{table}[h]
\caption{AGNs with detected lag between the IR and optical or UV variations}

    \begin{tabular}{lccl}
\hline
Object   & $\Delta t$, lag   & Band   &  References\\
         &   days      & (1~-~2)         &\\
         & (1 from 2)   &                 &\\      
\hline
NGC4151 & $30\div 60$  & $L(UBV)$ &  Penston\\
         &             &          & et al., 1974\\
         & $18\pm 6$    & $K(U)$   &  Oknyanskij\\
         &              &          & 1993\\
         & $35\pm 8$    & $K(UBV)$ &  Oknyanskij\\
         & $97\pm 10$   & $L(UBV)$ &  et al.,1999\\
         & $8\pm 4$     & $H(UBV)$ &  \\
         & $\sim 6$     & $J(UBV)$ &  \\
NGC3786  & $32\pm 7$    & $K(V)$   &  Nelson,1996)\\
NGC3783  & $\sim 78$    & $K(U)$   &  Glass,1992\\
F9       & $410\pm 110$ & $L(UV)$  &  Clavel \\
         & $385\pm 100$ & $K(UV)$  & et al.,1989\\
         & $250\pm 100$ & $H(UV)$  &\\
         & $-20\pm 100$ & $J(UV)$  &\\
NGC7469  & $ - $    &JHLK& Glass,1998\\
         & $0\pm 25$ & $J(U)$ & This paper\\
         & $41\pm 25$ & $K(U)$ &\\
         & $\ge 65$ & $L(U)$ &\\
         & $40\pm 20$ & $K(J)$ &\\
         & $73\pm 15$ & $L(J)$ &\\
         & $30\pm 20$ & $H(J)$ &\\
GQ Comae & $~\sim 250 $ & $K(UV)$ & Sitko\\
         & $~\sim 700 $ & $L(UV)$ &et al.,1993\\
         & $ 260\pm20 $ & $K(V)$  & This paper\\
         & $ 750\pm20 $ & $L(V)$  &\\
\hline
    \end{tabular}
\end{table}
}

{\bf 3. Observational data on the Optical-to-NIR time
delays in AGNs}\\[1mm]

By now, the time delay between optical
(UV) and NIR variations has been detected in several AGNs. The data on these
objects (including our results) are given in the Table 1.
The data which are not quite reliable (for example, results
for IIIZw2 (Lebofsky and Reike 1980) and  NGC 1566 (Baribaud et al. 1992) were not
included in the table. The objects where NIR radiation has nonthermal origin
(BLACs) and objects with a peculiar orientation, presence of superluminal
radio components (for example, 3C273) were not considered in the
paper too.\\[2mm]

{\bf 4. Estimation of $H_0$}\\[1mm]
Observed data are very good following to the theoretical relation (3)
for $H_0=50~km~s^{-1}~Mpc^{-1}$ (see Fig.1). 

\begin{figure}
  \includegraphics[width=\columnwidth]{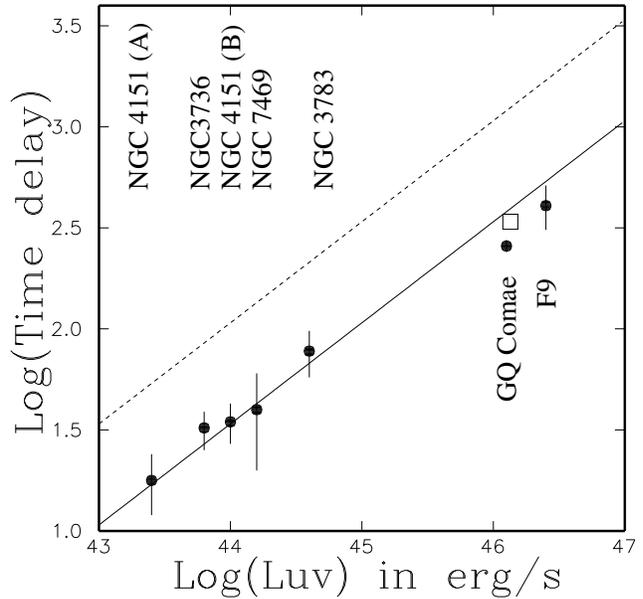}
  \caption[]{Dependence  luminosity -- time delay in logarithmic scale. Line
corresponds to the theoretical dependence (3), dashed line - to the (1). Points
correspond to the time delay data (for filter $K$) from  Table 1 and
UV luminosities estimated for $H_0=50~km~s^{-1}~Mpc^{-1}$ from the observed
fluxes. A box corresponds to the time delay for GQ Comae corrected for the
red shift (z=0.165).}
\end{figure}

So using (2) we have got the estimation
\begin{equation}
 H_0 \sim ~50~km~s^{-1}~Mpc^{-1}. \\[2mm]
\end{equation}
\linebreak\vfill\pagebreak\noindent
\\[2mm]

{\bf  Summary}\\[1mm]

We have combined published data on the optical-to NIR time delay
in AGNs.

We have made cross-correlation analysis of published data using
own code and have found the new values of time delays for 
NGC4151, 7469, GQ Comae.

We show that the observed time delays allow us to derive an
estimate of the Hubble constant value, however it is model 
depended.

The results presented here will be used as the groundwork
for  more detailed paper which are in preparation.
\\[2mm]

{\it Acknowledgements.} The author are thankful to Prof. K.Horne for
useful discussions.
\\[3mm]
\indent

{\bf References\\[2mm]}
Baribaud, T., Aloin, D., Glass, I.\ S., Pealat, D.: \h 1992, {\it Astron. Astrophys.,} 
 {\bf 256}, 375.\\
Barvainis, R.: 1987, {\it Ap.J.,} {\bf 320},  537.\\
Barvainis, R.: 1992, {\it Ap.J.,} {\bf 400},  502.\\
Clavel, J., Wamsteker, W., Glass, I.\ S.: 1989, {\it Ap.J.,} \h {\bf 337}, 236.\\
Collier S., Horn K., Wanders I., Peterson B.\ M.: 1999, \h {\it MNRAS,} {\bf 302}, L24.\\
Glass, I.\ S.: 1992, {\it MNRAS,} {\bf 256},  23P.\\
Glass, I.\ S.: 1998, {\it MNRAS,} {\bf 297},  18.\\
Goodwin, S.\ P., Gribbin, J., Hendry, M.\ A.: 1997, \h {\it A.J.,} {\bf 114}, 2212\\ 
Freedman W. et al.: 1989, {\it Ap.J.Suppl.Ser.,} {\bf 69}, 763.\\
Lebofsky, M.\ J., Reike, G.\ H.: 1980 {\it Nature} \h {\bf284}, 410.\\
Miyoshi, M. et al.: 1995 {\it Nature,} {\bf 373}, 127.\\
Nelson, B.\ O.: 1996, {\it Ap.J.,} {\bf 465},  87.\\
Oknyanskij, V.\ O.:1994, {\it Astron. Lett.,}  {\bf 19},  416.\\
Oknyanskij, V.\ O.: 1999, {\it Astron. Lett.,} {\bf 25},  483.\\
Penston, M.\ V., Balonek, T.\ J., Barker, E.\ S. et al.: \h 1974, {\it M.N.R.A.S,} {\bf 159},
 357.\\
Refsdal S.: 1964 {\it MNRAS,} {\bf 128}, 295.\\
Sandage, A. et al.: 1996, {\it Ap.J.,} {\bf 460},  L15.\\
Sanders D.\ B. et al.: 1989, {it Ap.J.,} {\bf 347}, 29.\\
Sitko, M.\ L., Sitko, A.\ K., Siemiginowska, A., \h Szczerba, R.: 1993,  {\it Ap.J.,} {\bf 409},
 139.\\
Syunyaev, R.\ A., Zel'dovich, Y.\ B.: 1980, {\it Ann. Rev. \h Astron. Ap.,} {\bf 18},  537.\\

\end{document}